\documentclass{article}
\begin{document}
\newcommand{\be}{\begin{equation}}
\newcommand{\ee}{\end{equation}}
\newcommand{\bea}{\begin{eqnarray}}
\newcommand{\eea}{\end{eqnarray}}
\title{Gravitational Acceleration of Spinning Bodies\\ From Lunar Laser Ranging Measurements}
\author{Kenneth Nordtvedt\\Northwest Analysis, 118 Sourdough Ridge Road, Bozeman MT 59715 USA\\{\it kennordtvedt@imt.net}}
\maketitle
\begin{center}
\section*{Abstract}
\end{center}
\begin{quote}
The Sun's $1/c^2$ order gravitational gradient accelerations of Earth and Moon, dependent on the motions of the latter bodies, act upon the system's internal angular momentum. Not only does this gravitational {\it spin-orbit force} (which plays an important part in calculating gravity wave signal templates from astrophysical sources) slightly accelerate the Earth-Moon system as a whole, it more robustly perturbs the internal Earth-Moon dynamics with a $5\;cm$ amplitude, synodically oscillating range contribution which is presently measured to $4\;mm$ precision by more than three decades of accumulated lunar laser ranging data.
\end{quote}

\section{Introduction}

The Earth-Moon system viewed as a whole is a {\it spinning body} moving through the gravitational field of the Sun, with the Moon's orbital motion around the Earth contributing five times as much to that system's total spin angular momentum as does the Earth's rotation about its axis. In metric theories of gravity, bodies with internal angular momentum follow trajectories in external gravitational fields which differ from those of so-called test bodies (and simple real bodies); but this difference of trajectories due to a gravitational {\it spin-orbit} acceleration has not yet been confirmed by observation.  The magnitude of the spin-orbit acceleration of the Earth-Moon system by the Sun is expected to be about $3\,\Omega^2SV/mc^2$, and to be radial and repulsive, with $\Omega^2=GM_s/R^3$ being the strength of the Sun's gravitational gradient, V being the orbital speed of the system around the Sun, $S$ and $m$ being the total internal angular momentum and total mass, respectively, of the system. The factor $3$ in this expression becomes $\gamma+2$ for scalar-tensor theories of gravity, with {\it Eddington} parameter $\gamma$ being less than one and $1-\gamma$ being a measure of the fractional participation of the scalar interaction in metric gravity.  And even this generalized strength parameter is dependent on the technical details of defining the spinning body's "center" chosen to express its trajectory (This issue will be further discussed later). 

This spin-orbit acceleration is a consequence of  underlying relativistic gravitational gradient forces (accelerations) which the Sun exerts separately on the Earth and Moon; these unequal accelerations thereby affect the lunar orbit relative to Earth.  The relativistic tidal force acts primarily on the Moon whose internal orbital angular momentum is about $80$ times that of the relatively immobile Earth.  The lunar orbit's task is to transmit the great bulk of this force to the Earth, and it does so by becoming polarized in the solar direction. Over 30 years of lunar laser ranging (LLR) has produced data which permits measurement of this orbital polarization with realistic precision of $4\;mm$ and with formal error of measurement already below the $1\;mm$ level \cite{diwm}; and the continuation of the LLR mission with upgraded technology and improved modeling should push this precision of fit for the lunar orbit to even substantially better levels \cite{apol}.  The characteristic size of the Moon's acceleration relative to Earth due to the Sun's relativistic gravitational gradient is $3\,ru\,V\Omega^2/c^2$, $ru$ being the Moon's angular momentum per unit mass.  This amounts to the fraction $2.6\;10^{-12}$ of the Sun's Newtonian acceleration, and the frequency of this perturbation on the lunar orbit is synodic.  The orbital response is the same as would occur due to an Equivalence Principle violation (unequal acceleration of Earth and Moon by the Sun) --- a polarization of the orbit along the solar direction \cite{nor68} \cite{nor94}, and with length of several centimeters which is an order of magnitude larger than the existing observational precision of that polarization's measurement.

\section{The Formal Acceleration of a Spinning Body} 

The expected governing tensor equation of motion for a spinning body of mass $m$ and spin tensor $J^{\alpha\beta}$) is the {\it Papapetrou equation} \cite{pape} 
\be
\frac{D}{d\tau}\left(m\,V^{\alpha}\;+\;V_{\sigma}\frac{D\,J^{\alpha\sigma}}{d\tau}\right)\;=\;-\,\frac{1}{2}\,R^{\alpha}_{\sigma\mu\nu}\,V^{\sigma}J^{\mu\nu}
\ee
whose right hand side will contain further couplings to gradients of the metric field's curvature tensor if the body's structure is extended and non-spherical.  In Equation (1) $m$ is the body's rest mass, $D/d\tau$ indicates tensor derivative with respect to proper time, $V^{\mu}=dx^{\mu}/d\tau$ is the body's four-velocity $dx^{\mu}/d\tau$, $R^{\alpha}_{\beta\mu\nu}$ is the metric tensor field's curvature tensor, and with supplementary conditions yet to be specified for the spin tensor.  This is not a fundamental equation of theory; it must rest on and be derivable from the basic gravitational interaction between elements of matter (contributions to a body's total internal angular momentum from integer or half-integer $\hbar$ quantum mechanical spin of elementary particles are an exception; fundamental theory must supply their coupling to gravity). 

A spinning body's motion through a metric gravitational field deviates from the geodesic trajectories of ordinary test bodies because of the coupling of the body's internal angular momentum to the curvature tensor which is composed of both first and {\it second} derivatives of the gravitational metric potentials; therefore no fundamental violation of universality of free fall is implied.  The leading order acceleration of a body from Equation (1), the Papapetrou equation, is 
\[
\delta\, A^i\;\cong\;\frac{1}{2mc^2}\left(R^i_{jkl}\,V^j\,J^{kl}\;+\;2\,R^i_{ook}\,J^{ok}\right)
\]
$i,\; j,\; k,\;...$ are spatial indices, $o$ is the time index. (Otherwise in this paper, Greek indices $\alpha,\;\beta,\;...$ range over all four space and time indices.)  Common indices are summed over their range. Assuming that the antisymmetric spin tensor $J^{\mu\nu}$ is orthogonal to the body's 4-velocity vector $J^{\mu\nu}V_{\nu}=0$, and that its purely spatial components are given by the body's internal angular momentum 
\[
J^{kl}\;=\;\epsilon^{klj}\,S_j\hspace{.5in}with\;S_j\;=\;\int \rho(\vec{r})\,(\vec{r}\times\vec{v})_j\,d^3r
\]
$\epsilon^{klj}$ being the completely antisymmetric spatial permutation tensor,  $\epsilon^{xyz}=1$, then the spin tensor's remaining components are
\[
-J^{ok}\;=\;J^{ko}\;\cong\;(\vec{V}\times\vec{S})^k/c
\]
In a quasi-static environment of outside bodies, the dominant metric potentials have the form 
\bea
g_{oo}\;&\cong&\;1\,-\,2\,U/c^2 \nonumber \\
-\,g_{ij}\;&\cong&\;\delta_{ij}\,(1\,+\,2\gamma\,U/c^2) \nonumber
\eea
$U$ is the outside world's Newtonian gravitational potential, and $\gamma$ is one of the PPN {\it Eddington} parameters which has value one in General Relativity but lesser value in scalar-tensor theories, for example.  With $\vec{g}\equiv\vec{\nabla}U$, the curvature tensor components needed above are 
\bea
R_{iook}\;&\cong&\;-\,[\vec{\nabla}\vec{g}\,]_{ik} \nonumber \nonumber \\
R_{ijkl}\;&\cong&\;-\,\gamma\,\left(\delta_{il}\,[\vec{\nabla}\vec{g}\,]_{jk}\;+\;\delta_{jk}\,[\vec{\nabla}\vec{g}\,]_{il}\;-\;\delta_{ik}\,[\vec{\nabla}\vec{g}\,]_{jl}\;-\;\delta_{jl}\,[\vec{\nabla}\vec{g}\,]_{ik}\right) \nonumber
\eea
$[\vec{\nabla}\vec{g}\,]$ is the $3\times 3$ spatial gravitational gradient tensor of the outside world. The spin-induced acceleration can then be written in the form
\[
\vec{g}_{so}\;\cong\;\frac{1}{m\,c^2}\;\left((\gamma+1)\,[\vec{\nabla}\vec{g}\,]\cdot(\vec{V}\times\vec{S})\;+\;\gamma\,([\vec{\nabla}\vec{g}\,]\cdot\vec{V})\times\vec{S}\right)
\]
with $\vec{V}$ being the velocity of the spinning system through the outside world's gravitational field.  Applying this to motion in the Sun's central gravity 
\[
[\vec{\nabla}\vec{g}\,]\;=\;\Omega^2\,[3\,\hat{R}\hat{R}-{\bf I}]
\]
with $\Omega^2=GM_s/R^3$, ${\bf I}$ being the identity matrix, and $\hat{R}$ being the unit vector toward the Sun.  Neglecting the inclination of the lunar orbit plane from the ecliptic, and assuming uniform motion around the Sun, the spin-induced acceleration is radial and of strength
\[
g_{so}\;\cong\;(\gamma+2)\,\frac{S\,V}{m\,c^2}\,\Omega^2\hspace{.5in}S\cong \frac{(m-m')m'}{m}\,ru
\]
with $m'$ being the Moon's mass, while $r$ and $u$ are the Earth-Moon separation and relative speeds, respectively. As a fraction of the total solar acceleration of the Earth-Moon system, this spin-induced acceleration is of size $3\,10^{-14}$ and directly unobservable in the foreseeable future.  

\section{Relativistic Tidal Force Effects on Motions of Earth and Moon}

Lunar laser ranging (LLR) is here examined as a method for measuring the effect of internal angular momentum on the Earth-Moon system's gravitational dynamics; the levels of precision recently reached in the on-going LLR mission make this now within reach \cite{diwm}. From the fundamental equations of motion for the Moon's coordinate position $\vec{r}(t)_1$ and the Earth's position $\vec{r}(t)_2$, a "center" coordinate for describing the system's dynamics $\vec{R}$ and another internal coordinate for giving the Moon-Earth separation $\vec{r}$ can be formed
\bea
\vec{R}\;&=&\;\frac{m_1\,\left(1+\left(\vec{V}\cdot\vec{u}_1-\frac{1}{2}\vec{g}\cdot\vec{x}_1\right)/c^2\right)\,\vec{r}_1\,+\,m_2\,\left(1+\left(\vec{V}\cdot\vec{u}_2-\frac{1}{2}\vec{g}\cdot\vec{x}_2\right)/c^2\right)\,\vec{r}_2}{m_1\,+\,m_2} \\
\vec{r}\;&=&\;\vec{r}_1\,-\,\vec{r}_2
\eea
with $\vec{g}$ being the solar acceleration at the chosen system center $\vec{R}$.  The inverse of these coordinate definitions are also useful
\bea
\vec{r}_1\;&=&\;\vec{R}\,+\,\frac{m_2}{m}\left(1+\left(\vec{V}\cdot\vec{u}_2-\frac{1}{2}\vec{g}\cdot\vec{x}_2\right)/c^2\right)\,\vec{r}\;\equiv\;\vec{R}\,+\,\vec{x}_1 \\ 
\vec{r}_2\;&=&\;\vec{R}\,-\,\frac{m_1}{m}\left(1+\left(\vec{V}\cdot\vec{u}_1-\frac{1}{2}\vec{g}\cdot\vec{x}_1\right)/c^2\right)\,\vec{r}\;\equiv\;\vec{R}\,+\,\vec{x}_2
\eea
with $m=m_1+m_2$.  These relationships reveal that the total Newtonian gravitational gradient force on the system contributes at relativistic order because of the chosen system center
\be
m_1\,\vec{x}_1\;+\;m_2\,\vec{x}_2\;=\;-\,\frac{m_1m_2}{mc^2}\,\vec{V}\cdot\vec{u}\,\vec{r}
\ee

An internal coordinate $\vec{\rho}$ which shows simpler dynamics more closely related to the observable range between Earth and Moon is constructed
\be
\vec{r}\;=\;\vec{\rho}\;-\;\frac{m_2-m_1}{mc^2}\left(\vec{V}\cdot\vec{\rho}\,\vec{\upsilon}\;+\;\gamma\,\vec{g}\cdot\vec{\rho}\,\vec{\rho}\;-\;\gamma\,\vec{g}\,\rho^2/2\right)
\ee
which generates pertinent differences between these two coordinate accelerations 
\be
\frac{d^2\vec{\rho}}{dt^2}\;=\;\frac{d^2\vec{r}}{dt^2}\;+\;\frac{m_2-m_1}{mc^2}\left(\vec{V}\cdot\vec{a}\,\vec{\upsilon}\;+\;2\vec{V}\cdot\vec{\upsilon}\,\vec{a}\;+\;(\vec{V}\cdot\vec{\nabla}\vec{g})\cdot\vec{\rho}\,\vec{\upsilon}\;+\;2\gamma(\vec{V}\cdot\vec{\nabla}\vec{g})\cdot(\vec{\rho}\vec{\upsilon}+\vec{\upsilon}\vec{\rho})\right)
\ee
with $d\vec{g}/dt=\vec{V}\cdot\vec{\nabla}\vec{g}$ for a static gravitational field; $\vec{\upsilon}=d\vec{\rho}/dt$ and $\vec{a}=d\vec{\upsilon}/dt$.

While the internal coordinate $\vec{\rho}$ is directly related to the LLR observations, the "center" coordinate is primarily useful in an analytic sense and is not observed at any instant, although its longer term evolution is a useful first approximation of where the Earth-Moon system is located in the solar system. A sub-system of the total system of gravitationally interacting bodies does not have a center of mass-energy because of its interaction energy with the total system, so the particular choice for the system's "center" coordinate, Equation (2), is motivated by the expression in the Appendix, Equation (xxxxxxx), for the conserved center-of-energy coordinate for an entire and isolated system of gravitationally interacting bodies.  Because the two bodies of the the Earth-Moon system are held together by the relatively elastic gravitational force, itself; LLR readily observes the internal perturbations produced by the basic post-Newtonian gravitational gradient forces which are responsible for the system's spin-related dynamics. 

If the Sun's acceleration of the Moon from Equation (xxxxxxxx) is expanded to the first tidal order, with the Newtonian and $1/c^2$ order motional corrections linear in both $\vec{V}$ and internal velocity $\vec{\upsilon}_1$ collected, there results  
\bea
(\vec{g}_1)_{tidal}\;&=&\;\Omega^2\;\left((3\hat{R}\,\hat{R}\cdot\vec{x}_1-\vec{x}_1\right) \nonumber \\
&+&\;2\left(\frac{m_2}{m}\right)^2\,\frac{\Omega^2}{c^2}\left(\gamma\,(3\hat{R}\,\hat{R}\cdot\vec{\rho}-\vec{\rho})\,\vec{V}\cdot\vec{\upsilon}\;-\;(1+\gamma)\,(3\hat{R}\,\hat{R}\cdot\vec{\rho}-\vec{\rho})\cdot(\vec{V}\,\vec{\upsilon}+\vec{\upsilon}\vec{V})\right)
\eea
and the analgous term for the Earth.  In the second line of Equation (9), which is already order $1/c^2$, the Newtonian approximation $\vec{x}_1\cong \vec{\rho}\,m_2/m$ is used, but $\vec{x}_1$ is retained on the first line because of the relationship given in Equation (6). 

\subsection*{Dynamics of the Earth-Moon System Center}

A combination of the tidal accelerations useful for evaluating the expression in Equation (2) is
\bea
\vec{g}_{tidal}\;&=&\;\frac{m_1\,(\vec{g}_1)_{tidal}\;+\;m_2\,(\vec{g}_2)_{tidal}}{m} \\
&=&\;\frac{m_1m_2}{m^2c^2}\Omega^2\left((2\gamma-1)\,\vec{V}\cdot\vec{u}\,(3\hat{R}\hat{R}\cdot\vec{r}-\vec{r})\,-\,(2\gamma+2)(3\hat{R}\hat{R}\cdot\vec{r}-\vec{r})\cdot(\vec{V}\vec{u}+\vec{u}\vec{V}\right)
\eea
The equation of motion for the defined "center" can then be assembled and altogether
\bea
\frac{d^2\vec{R}}{dt^2}\;&=&\;\vec{g}(\vec{R})\;+\;2\gamma\,\frac{m_1m_2}{m^2c^2}\,\left(\vec{V}\cdot\vec{u}\,[\vec{\nabla}\vec{g}\,]\cdot\vec{r}\;-\;\vec{r}\cdot[\vec{\nabla}\vec{g}\,]\cdot(\vec{V}\,\vec{u}+\vec{u}\,\vec{V})\right) \nonumber \\
&+&\;\frac{m_1m_2}{m^2c^2}\,\left(\vec{V}\cdot[\vec{\nabla}\vec{g}\,]\cdot(\vec{u}\,\vec{r}\,-\,\vec{r}\,\vec{u})\,-\,2\vec{u}\cdot[\vec{\nabla}\vec{g}\,]\cdot\vec{r}\,\vec{V}\right)
\eea
Setting $\hat{u}=\hat{S}\times\hat{r}$ and averaging over the monthly motion, $\langle\hat{r}\hat{r}\rangle=([{\bf I}]-\hat{S}\hat{S})/2$, this center's equation of motion becomes
\be
\frac{d^2\vec{R}}{dt^2}\;=\;\vec{g}(\vec{R})\;+\;\frac{1}{mc^2}\left(\gamma\,(\vec{V}\times\vec{S})\cdot[\vec{\nabla}\vec{g}\,]\;+\;(1+\gamma)(\vec{V}\cdot[\vec{\nabla}\vec{g}\,])\times\vec{S}\right)
\ee

But it should be noted that the orbital centers of both Earth and Moon  do not coincide with the system's chosen "center" coordinate $\vec{R}$. From Equations (4,5) the locations of Earth and Moon are given by
\bea
\vec{r}_1\;&=&\;\vec{R}\;-\;\frac{\vec{V}\times\vec{S}}{mc^2}\;-\;\left(\frac{m_2}{mc}\right)^2\,\vec{V}\cdot\vec{\rho}\,\vec{\upsilon}\;+\;\frac{m_2}{m}\,\vec{\rho} \\
\vec{r}_2\;&=&\;\vec{R}\;-\;\frac{\vec{V}\times\vec{S}}{mc^2}\;-\;\left(\frac{m_1}{mc}\right)^2\,\vec{V}\cdot\vec{\rho}\,\vec{\upsilon}\;-\;\frac{m_1}{m}\,\vec{\rho}
\eea
Both bodies have a common and slowly varying shift of their orbits' positions in the solar system; and each orbit viewed from the solar system rest frame shows the longitudinal oscillation in position due to special relativity's time transformation between inertial frames;
\[
t\;=\;\vec{V}\cdot\vec{\rho}/c^2\;+\;\sqrt{1-V^2/c^2}\,t'
\]
which for a given solar system bary-centric time $t'$, gives the time $t$ for finding the interbody vector at $\vec{\rho}(t)$ when viewed in the Earth-Moon system's instantaneous rest frame.  Defining the actual "center" about which the Earth and Moon orbits move 
\[
\vec{R}_c\;\equiv\;\vec{R}\;-\;\frac{1}{mc^2}\vec{V}\times\vec{S}
\]
then requires the adjustments
\[
\frac{d^2\vec{R}}{dt^2}\;=\;\frac{d^2\vec{R}_c}{dt^2}\;+\;(\vec{V}\cdot[\vec{\nabla}\vec{g}\,])\times\vec{S}
\]
in which $d^2\vec{V}/dt^2=\vec{V}\times[\vec{\nabla}\vec{g}\,]$ is used, and
\[
\vec{g}(\vec{R})\;=\;\vec{g}(\vec{R}_c)\;+\;(\vec{V}\times\vec{S})\cdot[\vec{\nabla}\vec{g}\,]
\]
The reexpression of Equation (13) in terms of $\vec{R}_c$ instead of $\vec{R}$ then shows a spin-orbit acceleration in agreement with that which emerges from the Papapetrou equation.
\[
\frac{d^2\vec{R}_C}{dt^2}\;=\;\vec{g}(\vec{R}_C)\;+\;\vec{g}_{so}
\]
but with light now shed on what system "center" coordinate achieves that agreement.  For a given Earth-Moon system orbital frequency $\Omega$ about the Sun, the orbital radius is then given by
\[
(R_c)^3\;=\;\frac{GM_S}{\Omega^2}\,\left(1\;-\;(\gamma+2)\frac{1}{R_c}\frac{S\,V}{mc^2}\right)
\]
which indicates a spin-orbit force induced shift in radial location of both Earth and Moon in amount $(\gamma+2)SV/3mc^2\cong 1.6\;mm$.

\subsection*{Dynamics of the Earth-Moon separation}

The relativistic solar tidal accelerations of the Moon and Earth, Equation (9), also produces a perturbation of the Earth-Moon separation which is directly measured by LLR.  
\bea
(\vec{g}_1)_{tide}-(\vec{g}_2)_{tide}\,&=&\,\Omega^2\,(3\hat{R}\hat{R}\cdot\vec{r}-\vec{r}) \nonumber \\
&+&\,\frac{m_2-m_1}{mc^2}\,2\Omega^2\,\left(\gamma\,\vec{V}\cdot\vec{\upsilon}\,(3\hat{R}\hat{R}\cdot\vec{\rho}-\vec{\rho})\;-\;(\gamma+1)(3\hat{R}\hat{R}\cdot\vec{\rho}-\vec{\rho})\cdot(\vec{V}\vec{\upsilon}+\vec{\upsilon}\vec{V}\right)
\eea
which coupled with Equations (8) produces the net internal equation of motion
\bea
\frac{d^2\vec{\rho}}{dt^2}\;&=&\;-\,\frac{Gm}{\rho^3}\vec{\rho}\;+\;\vec{\rho}\cdot[\vec{\nabla}\vec{g}\,] \nonumber \\
&+&\;\frac{1}{c^2}\left((2\gamma+2)\,\left(\vec{V}\cdot\vec{\upsilon}\,\vec{\rho}\cdot[\vec{\nabla}\vec{g}\,]\,-\,\vec{r}\cdot[\vec{\nabla}\vec{g}\,]\cdot\vec{\upsilon}\,\vec{V}\right)\;+\;2\gamma\,\vec{V}\cdot[\vec{\nabla}\vec{g}\,]\cdot\vec{\upsilon}\,\vec{r}\right)
\eea
with $[\vec{\nabla}\vec{g}\,]=\Omega^2\,(3\hat{R}\hat{R}-{\bf I})$ being the tidal tensor of the Sun's gravity.

The Newtonian tidal acceleration in Equation (17) produces the classic lunar {\it variation} which was first discovered analytically by Newton using his  universal law of gravity.  This perturbation consists of an orbit elongation transverse to and a contraction parallel to the solar direction, resulting in a range oscillation with half month period.   
\be
\delta\,\vec{\rho}_{var}\;\cong\;\rho\,\frac{\Omega^2}{\omega^2}\left(-\,cos(2D)\,\hat{\rho}\;+\;\frac{11}{8}\,sin(2D)\,\hat{\upsilon}\right) 
\ee
with $D$ abeing the Moon's synodic phase (angle between Sun and Moon as seen from Earth).  The relativistic tidal perturbation given by the second line of Equation (17) perturbs the lunar orbit at the synodic frequency. Its projections onto the radial and tangential directions of the lunar orbit are  
\bea
\vec{g}_{tide}\cdot\hat{r}\;&=&\;(2\gamma\,+\,4)\,\left(\frac{m_2-m_1}{mc^2}\right)\,\Omega^2\,ruV\;cos(D)\nonumber \\
\vec{g}_{tide}\cdot\hat{u}\;&=&\;0 \nonumber
\eea
in agreement with a previous analysis. The resulting radial perturbation in the Earth-Moon range is
\[
\delta r(t)\;\cong\;\;(\gamma+2)\,\left(\frac{m_2-m_1}{m}\right)\left(\frac{1-2\Omega/\omega+...}{1-7\Omega/\omega+...}\right)\, \frac{V\Omega\,r^2}{c^2}\,cos(D)\;\cong\;(6\;cm)\;cos(D)
\]
with the amplifying factor $1/(1-7\Omega/\omega)$ which almost doubles the perturbation being due to an unusual feedback mechanism between this perturbation of synodic frequency and the existing {\it variation} disturbance of the orbit at twice the synodic frequency \cite{nor94}.  This $5\;cm$ perturbation is well above the present precision, $4\;mm$, in measuring the amplitude of the Earth-Moon synodic range oscillation \cite{wilm}\cite  {di94}.  This perturbation is, of course, part of the outcome which occurs when the equations of motion given by Equation () are computer-integrated and the resulting coordinate positions of Earth and Moon are used to determine round trip ranging times by light signals.  The purpose here has been to make this internal relativistic tidal perturbation explicit and show its direct connection with the spin-orbit acceleration of a system as given by Equation ().  Without this $5\;cm$ amplitude contribution to the Earth-Moon range, the ranging data could not be successfully fit by the underlying theoretical model.

\clearpage
\begin{center}
\section*{Appendix}
\end{center}

In scalar-tensor metric theories of gravity, the first post-Newtonian ($1/c^2$ order) gravitational equations of motion for bodies $i,\;j,\;k\,=\,1,\,... N$ are based upon the lagrangian 
\bea
L\;&=&\;\sum_i\,m_i\,\left(-c^2\;+\; \frac{1}{2} \,v_i^2\;+\;\frac{1}{8c^2}\,v_i^4\right)\;+\;\frac{G}{2}\sum_{i,\,j\neq i}\frac{m_i\,m_j}{r_{ij}}\left(1\,-\,\frac{1}{2c^2}\left(\vec{v}_i\cdot\vec{v}_j\,+\,\vec{v}_i\cdot\hat{r}_{ij}\,\hat{r}_{ij}\cdot\vec{v}_j\right)\right)      \nonumber \\ 
&+&\;(2\gamma+1)\frac{G}{4c^2}\sum_{i,\,j\neq i}\frac{m_i\,m_j}{r_{ij}}(\vec{v}_i-\vec{v}_j)^2\;-\;(2\beta-1)\,\frac{G^2}{2c^2}\sum_{i,\,j\neq i,\,k\neq i}\frac{m_i\,m_j\,m_k}{r_{ij}\,r_{ik}}  
\eea
and the application of a least action principle which produces the dynamical equations
\[
\frac{d}{dt}\frac{\partial L}{\partial\vec{v}_a}\;=\;\frac{\partial L}{\partial\vec{r}_a}\hspace{.5in}a= 1,\, ...,\, N
\]
The form of the lagrangian given above does not take gravitational self-energy effects within the bodies into account; but that lagrangian, or the equations of motion, themselves, can be applied to modify the N-body equations for such self-gravity effects.  $\gamma$ and $\beta$ are two {\it Eddington} parameters which are each equal to one in pure tensor General Relativity but which generally differ in scalar-tensor gravity \cite{nosc}. The resulting equations of motion are \cite{diwm}
\bea
\vec{a}_i\;&=&\;\vec{g}_i\;+\;\sum_{j\neq i}\vec{g}_{ij}\,\left(\gamma\,\frac{1}{c^2}(\vec{v}_i-\vec{v}_j)^2\,-\,\frac{2}{c^2}\vec{v}_i\cdot\vec{v}_j\,+\,\frac{1}{c^2}v_j^2\,-\,\frac{3}{2c^2}(\hat{r}_{ij}\cdot\vec{v}_j)^2\right) \nonumber \\
&-&\;\sum_{j\neq i}\vec{g}_{ij}\cdot\left(\frac{2+2\gamma}{c^2}(\vec{v}_i-\vec{v}_j)\,+\,\frac{1}{c^2}\vec{v}_j\right)\,(\vec{v}_i-\vec{v}_j) \nonumber \\
&-&\;\frac{2\beta+2\gamma}{c^2}\,U_i\,\vec{g}_i\;-\;\sum_{j\neq i}\left(\vec{g}_{ij}\left(\frac{2\beta-1}{c^2}\,U_j\;-\;\frac{1}{2c^2}(\vec{r}_j-\vec{r}_i)\cdot\vec{g}_j\right)\;+\;\frac{3+4\gamma}{2c^2}\,U_{ij}\,\vec{g}_j\right)
\eea
with the underlying 2-body Newtonian potential and gravitational acceleration of $i$ due to $j$, and the cumulative potentials and gravitational accelerations being
\[
U_{ij}\;=\;\frac{G\,m_j}{r_{ij}}\hspace{.5in}\vec{g}_{ij}=\;\frac{G\,m_j}{r_{ij}^3}\,(\vec{r}_j-\vec{r}_i)\hspace{.5in}U_i\;=\;\sum_{j\neq i}U_{ij}\hspace{.5in}\vec{g}_i\;=\;\sum_{j\neq i}\vec{g}_{ij}
\]
and under the assumption that there are only gravitational accelerations acting among the bodies, $1/c^2$ contributions proportional to accelerations of bodies have been expressed with body accelerations replaced by the Newtonian gravity contributions, $\vec{a}_k\rightarrow\vec{g}_k$.  These equations of motion are computer-integrated by different analysis centers to produce the solar system ephemeris for its various bodies.  The resulting coordinate trajectories supplemented by metric gravity's expressions for the coordinate speed of light
\[
\frac{|d\vec{r}|}{dt}\;=\;c(\vec{r},\,t)\;=\;c_{\infty}\,\left(1-(1+\gamma)U(\vec{r},\,t)/c^2\right)
\]
and for the intrinsic rate of generally located and moving clocks
\[
d\tau\;=\;dt\,\left(1-v^2/2c^2-U(\vec{r},\,t)/c^2\right)
\]
are the basic ingredients for modeling the radio and laser ranging data between solar system bodies to $1/c^2$ ($1/c^3$ in time) order of accuracy. $U(\vec{r},\,t)$ is the Newtonian gravitational potential at location $\vec{r}(t)$ of light ray or clock as produced by all solar system sources.

The equations of motion resulting from this Lagrangian fulfill the classic conservation laws for {\it total system} momentum, angular momentum, and energy
\[
\frac{d\vec{P}}{dt}\;=\;\frac{d\vec{L}}{dt}\;=\;\frac{dE}{dt}\;=\;0 
\]
with
\[
\vec{P}\;=\;\sum_i\frac{\partial L}{\partial \vec{v}_i}\hspace{.5in}\vec{L}\;=\;\sum_i\vec{v}_i\times\frac{\partial L}{\partial\vec{v}_i}\hspace{.5in}E\;=\;-L\,+\,\sum_i\vec{v}_i\cdot\frac{\partial L}{\partial\vec{v}_i}
\] 
And the total system possesses a {\it center of energy} which does not accelerate
\be
\vec{R}\;=\;\frac{1}{E}\,\sum_i\,m_i\,\vec{r}_i\,\left(c^2\,+\,\frac{1}{2}v_i^2\,-\,\frac{1}{2}\,U_i\right)\hspace{.5in}\frac{d^2\vec{R}}{dt^2}\;=\;0
\ee

Applying the complete N-body, first post-Newtonian order gravitational equations of motion given in by Equation (), to the Sun-Earth-Moon system, two useful combinations of $1/c^2$ order acceleration terms proportional to the Newtonian accelerations between Earth and Moon, then multiplied by both the Earth-Moon system velocity relative to the Sun $\vec{V}$ and the velocity of the Moon relative to Earth $\vec{u}$, are here displayed
\bea
\vec{a}_1\;-\;\vec{a}_2\;&=&\;\frac{G\,(m_2-m_1)}{c^2r^3}\,\left(2\,\vec{V}\cdot\vec{u}\,\vec{r}\;+\;\vec{V}\cdot\vec{r}\,\vec{u}\right) \\
\frac{m_1\vec{a}_1+m_2\vec{a}_2}{m_1+m_2}\;&=&\;\frac{G\,m_1m_2}{(m_1+m_2)c^2r^3}\,\left(2\,\vec{V}\cdot\vec{u}\,\vec{r}\,+\,2\,\vec{V}\cdot\vec{r}\,\vec{u}\;-\;3\,\vec{V}\cdot\hat{r}\,\vec{u}\cdot\hat{r}\,\vec{r}\right)
\eea
because in the body of this paper they play an important role in establishing the equations of motion for the center and internal coordinates, $\vec{R}$ and $\vec{\rho}$, respectively.

{\bf This work has been supported by the National Aeronautics and Space Administration Contract NASW - 00011.}

\end{document}